# Stress Constrained Thermo-Elastic Topology Optimization with Varying Temperature Fields via Augmented Topological Sensitivity Based Level-Set

Shiguang Deng, Krishnan Suresh *

Mechanical Engineering, University of Wisconsin, Madison, USA

## Abstract

Engineering structures must often be designed to resist thermally induced stresses. Significant progress has been made on the design of such structures through thermo-elastic topology optimization. However, a computationally efficient framework to handle stress-constrained large-scale problems is lacking. The main contribution of this paper is to address this limitation. In particular, a unified topological-sensitivity (TS) based level-set approach is presented in this paper for optimizing thermo-elastic structures subject to non-uniform temperatures. The TS fields for various thermo-elastic objectives are derived, and, to address multiple constraints, an augmented Lagrangian method is developed to explore Pareto topologies. Numerical examples demonstrate the capability of the proposed framework to solve large-scale design problems. Comparison is made between pure elastic problems, and its thermo-elastic counterpart, shedding light on the influence of thermo-elastic coupling on optimized topologies.

## 1. Introduction

Engineering structures must often be designed to resist thermally and mechanically induced stresses. As an illustrative example, consider the support structure for a combustion-exhaust system of a high Mach supersonic airplane in Figure 1. During flight, cold air is sucked into inlet, mixed with fuel and ignited within the combustion chambers, with temperature climbing up to 1000 °C, then remixed with cool air and expelled. The support structures are therefore subject significant thermal gradients, giving rise to thermally induced stresses. In addition, the support structures are also subject to mechanical loads from aerodynamic effects, exhaust flow and pressures from adjoining airframe structures. The topology optimization of such support structures is the main focus of this paper.

Topology optimization has rapidly evolved from an academic exercise into an exciting discipline with numerous industrial applications [1], [2]. Applications include optimization of aircraft components [3], [4], spacecraft modules [5], automobiles components [6], cast components [7], compliant mechanisms [8]–[11], etc. Popular methods for thermo-elastic topology optimization include: homogenization, Solid Isotropic Material with Penalization (SIMP), Rational Approximation of Material Properties (RAMP), evolutionary structural optimization (ESO) and level-set.

Under the category of homogenization, a pioneering method [12] was developed to minimize structural compliance subject to a volume constraint where 2D thermo-elastic structures were represented by micro-void models. Thermal gradient was shown to have a significantly different impact on final topologies compared to uniform temperature field. Sigmund and Torquato [13] designed composites with extremal thermal expansion coefficients by using a three-phase topology optimization method which was proposed to find the distribution of each material phase by optimizing thermo-elastic properties subject to constraints of elastic symmetry and volume constraint. Jog [14] extended thermo-elastic compliance function to a nonlinear case, which was then minimized by using a density based linear penalization method and a perimeter constraint for regularization.

Over the last two decades, SIMP has evolved into one of the most popular topology optimization methods, due to its simplicity and generality, with applications ranging from fluids, structural mechanics, and multi-physics optimization [15]. Many pioneering researchers have applied SIMP to solve thermo-elastic design problems as well. In [16], structural strain energy was minimized after considering thermo-mechanical coupling where the sensitivity was calculated via an adjoint approach and the optimization was solved using the method of moving asymptotes. Deaton and Grandhi [17] presented a design scenario for restrained thin shell structure in a homogeneous thermal environment where the traditional design approach by accommodating thermal expansion to eliminate thermal stress was impossible. SIMP was set up and it was shown a typical compliance minimization in the presence of thermal loading did not guarantee favorable designs. Yang and Li [18] minimized structural dynamic compliance at resonance frequencies in a thermal environment, leading to the conclusion that final topologies were strongly affected by excited modes and load locations. In [19], Liu compared two thermo-elastic TO formulations: volume constrained compliance minimization and weight minimization with displacement constraint where the influence of different SIMP penalty factors on thermal and mechanical fields were studied. Zhang [20] investigated the difference between two minimization objectives in thermo-elastic TO formulations: mean compliance and elastic strain energy through sensitivity analysis. A concept of load sensitivity was introduced to interpret quantitatively the influence of thermal and mechanical loads on final topologies. Chen [21] presented a unified TO algorithm for multi-functional 3D finite periodic structures where the sensitivity at the corresponding locations at different components are regulated to maintain structural periodicity. Multiple objectives were simultaneously optimized through a weighted average method where thermo-elastic coupling was also considered. Pedersen [22] questioned the usage of thermo-elastic compliance as an objective, and suggested an alternate formulation of minimizing the maximum von Mises stress. Although the solutions provided a sound theoretical foundation, one of the challenges with SIMP is that the material interpolation exhibits zero slope at zero density, leading to parasitic effects in thermo-elastic problems [23], [24].

---

* Corresponding author (ksuresh@wisc.edu).

By employing a slightly different material interpolation scheme, RAMP [23] was reported to successfully overcome this deficiency with its superior performance over SIMP demonstrated in [24]. Gao and Zhang [25] proposed to penalize thermal stress coefficient, which is the product of thermal expansion coefficient and Young's modulus, and RAMP was shown to be advantageous over SIMP. Pedersen and Pedersen [26] compared SIMP, RAMP and an alternate two parameter interpolation scheme where the influence of interpolation on compliance sensitivity analysis was studied and the sensitivity of local von Mises stress was derived for problems with a uniform temperature elevation.

Substantial progress has also been made in ESO [27] where elements are gradually removed from design domain based on their relative significance order, while a BESO [28] addresses some of the limitations of ESO by permitting the insertion of elements. For thermo-elastic TO, ESO was seen as one of the earliest approaches used for solving such design problems. Li adopted ESO to solve fully stressed thermo-elastic topology design problems and later extended for thermal displacement minimization [29]. In [30] ESO was utilized to achieve a multi-criterion design for structures in thermal environment where material usage efficiency was measured by thermal stress levels and heat flux density. Li [31] developed an ESO procedure for design cases with uniform, non-uniform and transient temperature fields subject to single and multiple heat loads. Relative elemental efficiency defined in terms of thermal stress levels was employed to achieve the highest efficacy for material usage.

The level-set method was developed in [32] and introduced to structural optimization in [33]. Its primary advantage over other TO methods is that the boundary is well defined at all times. In the discipline of thermo-elastic structural designs, the level-set method was first reported by Xia and Wang [34] where a structural mean compliance was minimized with volume constraints. Sensitivity analysis of continuum body was conducted with respect to free boundaries which were smoothened by a geometric energy team during optimization process. In [35], a level-set based framework was developed to study the effects of including material interface properties to thermo-elastic multi-phase structures. Finite material interpolations with monotonic and non-monotonic property variations were utilized to guarantee material properties continue change across interfaces. Deng and Suresh [36] exploited the topological sensitivity based level-set method to solve 2D stress constrained TO problems subject to homogeneous temperature change, which was later extended to solve buckling constrained problems in a thermal environment [37].

Despite these advances, two research gaps can be identified. First, an efficient framework to address large-scale 3D thermo-elastic topology optimization problems, with millions of degrees of freedom, has not been demonstrated. Second, most studies are limited to uniform temperature scenarios. In this paper, a new level-set method is proposed to address these two gaps. It combines a discrete approximation of the topological sensitivity with augmented Lagrangian formulation to solve spatially varying temperature problems subject to a variety of constraints. The sensitivity of thermally induced p-norm stress is derived, for the first time. Finally, to address the computational challenges, the assembly-free deflated finite element method proposed in [38] is extended here for efficient large-scale 3D thermo-elastic optimization.

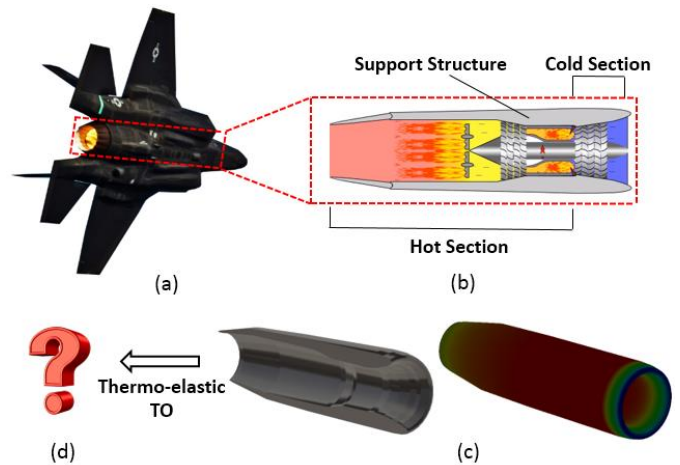

**Figure 1:** (a) Aero-gas turbine engine in an airplane; (b) Combustion-exhaust system; (c) Cross-section view and temperature distribution of the support structure; (d) Optimized support structure to withdraw thermal stress

## 2. Technical background

### 2.1 Thermo-elasticity

We restrict our attention here to weakly-coupled thermo-elastic problems, i.e., temperature influences displacements, but not the reverse. Finite element analysis (FEA) of such problems essentially reduces to posing and solving two linear systems of equations [39]. First, if the temperature field is spatially varying, one solves the thermal problem:

$$\mathbf{K}_t \mathbf{t} = \mathbf{q} \quad (2.1)$$

This is then followed by the structural problem:

$$\mathbf{K}\mathbf{d} = \mathbf{f}_{st} + \mathbf{f}_{th} \quad (2.2)$$

In the above two equations,

$\mathbf{t}$ : Temperature field

$\mathbf{d}$ : Displacement field

$\mathbf{K}_t$ : Thermal stiffness matrix

$\mathbf{K}$ : Structural stiffness matrix $\quad (2.3)$

$\mathbf{q}$ : Heat flux

$\mathbf{f}_{st}$ : Structural load

$\mathbf{f}_{th}$ : Thermal load

The thermal load vector in Equation (2.2) is formed by assembling the load for each finite element via [40]:

$$\mathbf{f_{th}} = \sum_{e=1}^{N} (\mathbf{f_e^{th}}) \qquad (2.4)$$

where

$$\mathbf{f_e^{th}} = \int_{\Omega_e} \mathbf{B_e^T D_e \varepsilon_e^{th}} d\Omega \qquad (2.5)$$

$$\mathbf{\varepsilon_e^{th}} = \alpha(t_e - t_0)\mathbf{\Phi}^T \qquad (2.6)$$

with

$\mathbf{f_e^{th}}$ : Nodal thermal load vector for each element

$\Omega_e$ : Element domain

$\mathbf{B_e}$ : Element strain-displacement matrix

$\mathbf{D_e}$ : Element elasticity matrix

$\mathbf{\varepsilon_e^{th}}$ : Element thermal strain vector

$\alpha$ : Thermal expansion coefficient  (2.7)

$t_e$ : Arithmetic average of nodal temperature fields in one element

$t_0$ : Reference temperature

$\mathbf{\Phi}$ : [1 1 1 0 0 0] in 3D; [1 1 0] in 2D

$e$ : Each finite element

$N$ : Total number of finite elements

Summing up contributions from all elements, the load due to thermal effects can be written as:

$$\mathbf{H}\Delta\mathbf{t} = \mathbf{f_{th}} \qquad (2.8)$$

where

$$\mathbf{H} = \sum_{e=1}^{N} (\int_{\Omega_e} \mathbf{B_e^T D_e} \alpha \mathbf{\Phi}^T d\Omega) \qquad (2.9)$$

$$\Delta\mathbf{t} = \mathbf{t} - \mathbf{t_0} \qquad (2.10)$$

where $\mathbf{t_0}$ is a reference temperature vector with the size of degrees of freedom.

The stresses are obtained by multiplying the material tensor with the difference between total strain and thermal strain [40]:

$$\mathbf{\sigma_e} = \mathbf{D_e} \left( \mathbf{B_e d_e} - \mathbf{\varepsilon_e^{th}} \right) \qquad (2.11)$$

The compliance for a thermo-elastic system is defined as:

$$J = (\mathbf{f_m} + \mathbf{f_{th}})^T \mathbf{d} = \mathbf{d}^T \mathbf{K d} \qquad (2.12)$$

**2.2 Thermo-elastic topology optimization**

Finding a suitable optimization objective in thermo-elastic TO is important. Using compliance as the design objective could lead to a "no structure" design. Instead, minimization of the maximum von Mises stress was argued as being more appropriate [41]. Further, it was shown in [20] that treating thermo-elastic compliance and elastic strain energy as objectives may lead to different topologies.

In this paper, the objective is to minimize volume, subject to both compliance and stress constraints, avoiding the scenario of "no structure" design. Further, both thermal and structural loads are considered throughout this paper.

A thermo-elastic TO problem can now be posed as:

$$\begin{aligned} &\underset{\Omega \subset \Psi}{Min}\, \varphi \\ &g_i(\mathbf{d}, \mathbf{t}, \Omega) \leq 0; i = 1, 2, ..., m \\ &\text{subject to} \\ &\mathbf{K_t t = q} \\ &\mathbf{Kd = f_{st} + f_{th}} \end{aligned} \qquad (2.13)$$

where:

$\varphi$ : Objective to be minimized

$\Omega$ : Topology to be computed

$\Psi$ : Design domain  (2.14)

$g_i$ : Constraints

$m$ : Number of constraints

In words, the objective is to find an optimal material layout ($\Omega$) within the design domain ($\Psi$) such that the quantity of interest ($\varphi$) is minimized while the constraints ($g_i$) are satisfied. Typical constraints include compliance, p-norm von Mises stress, buckling safety factor, etc.

A special case of Equation (2.13) is the volume minimization problem with volume, compliance and stress constraints:

$$\begin{aligned} &\underset{\Omega \subset \Psi}{Min}\, |\Omega| \\ &|\Omega| \geq v_f |\Psi| \\ &J \leq \alpha_1 J_0 \\ &\sigma \leq \alpha_2 \sigma_0 \\ &\text{subject to} \\ &\mathbf{Kd = f_m + f_{th}} \\ &\mathbf{K_t t = q} \end{aligned} \qquad (2.15)$$

As in any optimization problem, it will terminate if any of the following conditions are met: (1) volume fraction reaches $v_f$, or (2) compliance reaches $\alpha_1$ times the initial value, or (3) von Mises stress reaches $\alpha_2$ times the initial value.

Since compliance and p-norm stress are imposed as constraints, their sensitivity analysis is presented here. In addition, the sensitivity to displacement was also derived, and compared against the derivation in [42] for correctness.

## 3. Proposed method

In order to solve the above problem, we address sensitivity analysis in Section 3.1. Then, in Section 3.2, we discuss how one can directly use the sensitivity fields as a level-set to carry out topology optimization. Finally, in Section 3.3, we explicitly address the constraints through augmented Lagrangian formulation.

### 3.1 Sensitivity analysis

Let Q be any quantity of interest in a thermo-elastic optimization problem; Q can either be an objective, or a constraint. The following equations are derived to compute the sensitivity of Q with respect to a topological change for problems with spatially varying temperature fields.

The sensitivity of Q with respect to a topological design variable $\mathbf{x}$ is denoted by:

$$Q' \equiv \frac{\partial Q}{\partial \mathbf{x}} \tag{3.1}$$

The derivatives of the global stiffness matrix will be denoted by:

$$\mathbf{K'} \equiv \frac{\partial \mathbf{K}}{\partial \mathbf{x}} \tag{3.2}$$

By assuming the quantity of interest Q is a function of temperature $\mathbf{t}$ and displacement $\mathbf{d}$, its sensitivity filed can be expressed as:

$$Q' = (\nabla_\mathbf{d} Q)^T \mathbf{d'} + (\nabla_\mathbf{t} Q)^T \mathbf{t'} \tag{3.3}$$

By taking derivative of Equation (2.2) with respect to design variable $\mathbf{x}$, we obtain:

$$\mathbf{K'd} + \mathbf{Kd'} = \mathbf{f_{st}}' + \mathbf{f_{th}}' \tag{3.4}$$

If one assumes the external structural load $\mathbf{f}_{st}$ is independent of design variables, its sensitivity can be dropped from Equation (3.4):

$$\mathbf{d'} = \mathbf{K}^{-1}(\mathbf{f_{th}}' - \mathbf{K'd}) \tag{3.5}$$

On the other hand, since thermal load $\mathbf{f}_{th}$ depends on design variable $\mathbf{x}$, the change in thermal load due to a topological change can be calculated by taking derivative of Equation (2.8):

$$\mathbf{f_{th}}' = \mathbf{H'\Delta t} + \mathbf{H t'} \tag{3.6}$$

By substituting Equation (3.5) and (3.6) into Equation (3.3), we have:

$$Q' = \begin{bmatrix} (\nabla_\mathbf{d} Q)^T \mathbf{K}^{-1}(\mathbf{H'\Delta t} - \mathbf{K'd}) + \\ \left((\nabla_\mathbf{d} Q)^T \mathbf{K}^{-1}\mathbf{H} + (\nabla_\mathbf{t} Q)^T\right)\mathbf{t'} \end{bmatrix} \tag{3.7}$$

If one also assumes that the thermal flux in Equation (2.1) is independent of a topological change. Then, taking derivative of Equation (2.1) gives us:

$$\mathbf{t'} = -\mathbf{K}_t^{-1}\mathbf{K}_t'\mathbf{t} \tag{3.8}$$

Substituting Equation (3.8) into Equation (3.7):

$$Q' = \nabla_\mathbf{d} Q^T \mathbf{K}^{-1}(\mathbf{H'\Delta t} - \mathbf{K'd}) - \\ \left((\nabla_\mathbf{d} Q)^T \mathbf{K}^{-1}\mathbf{H} + (\nabla_\mathbf{t} Q)^T\right)(\mathbf{K}_t^{-1}\mathbf{K}_t'\mathbf{t}) \tag{3.9}$$

For ease of computation, two *adjoints* $\boldsymbol{\lambda}$ and $\boldsymbol{\omega}$ are introduced as follows:

$$\mathbf{K}\boldsymbol{\lambda} = -\nabla_\mathbf{d} Q \tag{3.10}$$

$$\mathbf{K}_t \boldsymbol{\omega} = (\mathbf{H}^T \boldsymbol{\lambda} - \nabla_\mathbf{t} Q) \tag{3.11}$$

Equation (3.9) can then be simplified as:

$$Q' = -\boldsymbol{\lambda}^T \mathbf{H'\Delta t} + \boldsymbol{\omega}^T \mathbf{K}_t'\mathbf{t} + \boldsymbol{\lambda}^T \mathbf{K'd} \tag{3.12}$$

For clarity, one can express Equation (3.12) as:

$$Q' = Q_{th}' - Q_{st}' \tag{3.13}$$

with

$$Q_{th}' = -\boldsymbol{\lambda}^T \mathbf{H'\Delta t} + \boldsymbol{\omega}^T \mathbf{K}_t'\mathbf{t}$$
$$Q_{st}' = -\boldsymbol{\lambda}^T \mathbf{K'd} \tag{3.14}$$

Observe that, as the topology evolves, the sensitivity in Equation (3.12) can take either a positive or a negative value. While this non-monotonic behavior can pose challenges for traditional monotonic approximation methods, it can be properly approximated with non-monotonous approaches like globally convergent version of MMA (GCMMA) and gradient-based MMA (GBMMA) [43]. In this paper, we employ topological sensitivity based level-set method which is proven to be robust and efficient, as illustrated later through numerical experiments. Further, observe that the adjoints $\boldsymbol{\lambda}$ and $\boldsymbol{\omega}$ depend on the quantity of interest $Q$. Three specific instances of $Q$ are considered below.

Displacement

If the quantity of interest is a displacement at a point $a$, that is:

$$Q \equiv d_a = \mathbf{1}^T \mathbf{d} \tag{3.15}$$

Then, the gradients defined in Equation (3.10) and (3.11) can be found as:

$$\nabla_\mathbf{d} Q = \frac{\partial(\mathbf{1}^T \mathbf{d})}{\partial \mathbf{d}} = \mathbf{1} \tag{3.16}$$

$$\nabla_\mathbf{t} Q = \frac{\partial(\mathbf{1}^T \mathbf{d})}{\partial \mathbf{t}} = \frac{\partial \mathbf{d}}{\partial \mathbf{t}}\mathbf{1} \tag{3.17}$$

In order to compute the term of $\nabla_\mathbf{t} \mathbf{d}$, we take derivative of Equation (2.2) with respect to temperature field $\mathbf{t}$:

$$\nabla_\mathbf{t} \mathbf{d} = \frac{\partial \mathbf{d}}{\partial \mathbf{t}} = \mathbf{K}^{-1}\frac{\partial(\mathbf{f_{th}} + \mathbf{f_{st}})}{\partial \mathbf{t}} \tag{3.18}$$

Since structural load $\mathbf{f}_{st}$ is independent of temperature field $\mathbf{t}$, we have:

$$\frac{\partial \mathbf{d}}{\partial \mathbf{t}} = \mathbf{K}^{-1}\frac{\partial \mathbf{f_{th}}}{\partial \mathbf{t}} = \mathbf{K}^{-1}\frac{\partial(\mathbf{H}(\mathbf{t} - \mathbf{t}_0))}{\partial \mathbf{t}} = \mathbf{K}^{-1}\mathbf{H} \tag{3.19}$$

Substituting Equation (3.19) back to Equation (3.17), we have:

$$\nabla_\mathbf{t} Q = \mathbf{K}^{-1}\mathbf{H1} \tag{3.20}$$

Then, substituting Equation (3.16) and (3.20) to Equation (3.10) and (3.11), the two adjoints can be expressed as:

$$\mathbf{K}\boldsymbol{\lambda} = -\mathbf{1} \tag{3.21}$$

$$\mathbf{K}_t \boldsymbol{\omega} = (\mathbf{H}^T \boldsymbol{\lambda} - \mathbf{K}^{-1}\mathbf{H1}) \tag{3.22}$$

Substituting Equation (3.21) and (3.22) back to Equation (3.12) leads to:

$$d_a' = \boldsymbol{\lambda}^T(\mathbf{K'd} - \mathbf{H'\Delta t}) + \boldsymbol{\omega}^T \mathbf{K}_t'\mathbf{t}$$
$$= (\mathbf{K}^{-1}\mathbf{1})^T(\mathbf{H'\Delta t} - \mathbf{K'd}) + (\mathbf{H}^T\mathbf{K}^{-1}\mathbf{1} + \mathbf{1}^T\mathbf{K}^{-1}\mathbf{H})\mathbf{t'} \tag{3.23}$$

Equation (3.23) is identical with the one in [44]. A detailed comparison between the two can be found in the Appendix.

Compliance

If the quantity of interest is compliance, the adjoints are given by:

$$\boldsymbol{\lambda} = -\mathbf{K}^{-1}\mathbf{f} = -\mathbf{d} \tag{3.24}$$

$$\mathbf{K}_t\boldsymbol{\omega} = -\mathbf{H}^T\mathbf{d} - \nabla_t Q \tag{3.25}$$

Therefore, after substituting Equation (3.24) and (3.25) back to Equation (3.12), the compliance sensitivity can be simplified to:

$$J' = \boldsymbol{\omega}^T \mathbf{K}_t'\mathbf{t} + \mathbf{d}^T \mathbf{H}'\Delta\mathbf{t} - \mathbf{d}^T\mathbf{K}'\mathbf{d} \tag{3.26}$$

<u>P-norm von Mises stress</u>

If the quantity of interest is p-norm stress, that is:

$$Q \equiv \left(\sum_e (\sigma_e)^p\right)^{1/p} \tag{3.27}$$

where:

$$\sigma_e = \frac{1}{\sqrt{2}}\sqrt{\begin{array}{l}(\sigma_{11}-\sigma_{22})^2 + (\sigma_{11}-\sigma_{33})^2 + ...\\ (\sigma_{22}-\sigma_{33})^2 + ...\\ 6(\sigma_{12}\sigma_{12} + \sigma_{13}\sigma_{13} + \sigma_{23}\sigma_{23})\end{array}} \tag{3.28}$$

Then, the adjoint $\boldsymbol{\lambda}$ defined in Equation (3.10) is given by

$$\mathbf{K}\boldsymbol{\lambda} = -\nabla_\mathbf{d} Q \equiv \mathbf{g} \tag{3.29}$$

where $\mathbf{g}$ defines the right-hand side of this adjoint problem:

$$\mathbf{g} = -\frac{1}{p}\left(\sum_e (\sigma_e)^p\right)^{\frac{1}{p}-1}\left[\sum_e \mathbf{g_e}\right] \tag{3.30}$$

where $\mathbf{g}$ is assembled from elemental vector $\mathbf{g}_e$ which is defined by:

$$\mathbf{g_e} = \frac{1}{\sqrt{2}} p(\sigma_e)^{p-2}\begin{pmatrix}(\sigma_{11}-\sigma_{22})(F_{1,:}-F_{2,:}) +\\ (\sigma_{11}-\sigma_{33})(F_{1,:}-F_{3,:}) +\\ (\sigma_{22}-\sigma_{33})(F_{2,:}-F_{3,:}) +\\ 6(\sigma_{12}F_{4,:}+\sigma_{13}F_{5,:}+\sigma_{23}F_{6,:})\end{pmatrix} \tag{3.31}$$

$$\mathbf{F} = \mathbf{D_e B_e} \tag{3.32}$$

where $\mathbf{B}_e$ is element strain-displacement matrix defined in Equation (2.7); please see [45] for details.

To account for the second term on the right hand side of Equation (3.11), we introduce another adjoint $\xi$ by:

$$\mathbf{K}_t\boldsymbol{\xi} = -\nabla_t Q \equiv \mathbf{g_t} \tag{3.33}$$

where $\mathbf{g_t}$ defines the right-hand side of this adjoint equation and it is assembled from the elemental vector $\mathbf{g}_{te}^a$:

$$\mathbf{g_t} = -\frac{1}{p}(\sum_e (\sigma_e)^p)^{\frac{1}{p}-1}\left[\sum_e \mathbf{g_{te}^a}\right] \tag{3.34}$$

With

$$\mathbf{g_{te}^a} = [g_{te}, \ g_{te}, \ g_{te}, \ g_{te}, \ g_{te}, \ g_{te}, \ g_{te}, \ g_{te}]^T \tag{3.35}$$

$$g_{te} = \frac{1}{\sqrt{2}}p(\sigma_e)^{p-2}\begin{pmatrix}(\sigma_{11}-\sigma_{22})(G_1-G_2)+\\ (\sigma_{11}-\sigma_{33})(G_1-G_3)+\\ (\sigma_{22}-\sigma_{33})(G_2-G_3)+\\ 6(\sigma_{12}G_4+\sigma_{13}G_5+\sigma_{23}G_6)\end{pmatrix} \tag{3.36}$$

where the components $G_i$ are defined via:

$$\mathbf{G} = \frac{1}{8}\alpha \mathbf{D_e \Phi} \tag{3.37}$$

where $\mathbf{D}_e$ and $\boldsymbol{\Phi}$ were defined in Equation (2.7).

Once the adjoints $\boldsymbol{\lambda}$ and $\xi$ are solved, they can be plugged into Equation (3.11) to compute the adjoint $\boldsymbol{\omega}$. From there, the sensitivity of the p-norm von Mises stress can be easily obtained. It should be noted that the sensitivity analysis for problems with uniform temperature change is just a special case of this derivation, which can be easily obtained by dropping the temperature variation term $\mathbf{t}'$ from Equation (3.3).

### 3.2 Level-set pareto tracing

A simple approach to exploiting the above sensitivity fields in topology optimization is to 'kill' mesh-elements with low values. However, this will lead to instability. Alternately, the sensitivity field can be used to introduce holes via an auxiliary level-set [46]. In this paper, we treat the sensitivity field as a level-set, as described next (also see [47]).

To illustrate, consider the pure structural problem illustrated in Figure 2a. For example, one can compute the elastic compliance sensitivity field from Equation (3.26) by setting the temperature change to 0. The resulting elastic compliance sensitivity field is illustrated in Figure 2 (b) where the field has been normalized.

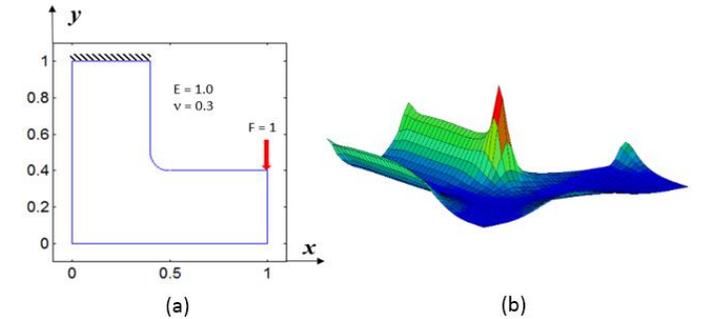

**Figure 2:** (a) A L-beam is subject to a tip load; (b) Its elastic compliance sensitivity field.

Now consider a cutting plane defined by a parameter $\tau$; one can define a domain $\Omega^\tau$ per:

$$\Omega^\tau = \{(x,y,z)\,|\,Q' > \tau\} \tag{3.38}$$

In other words, the domain $\Omega^\tau$ is the set of all points where the sensitivity field exceeds a given value $\tau$ [48]. The $\tau$ value is, in turn, computed from the volume fraction desired, leading to a simple binary search algorithm: find the value $Q'_{\min} \leq \tau \leq Q'_{\max}$ such that the resulting topology is of a desired volume fraction; for example, an induced domain with 95% volume fraction is illustrated in Figure 3.

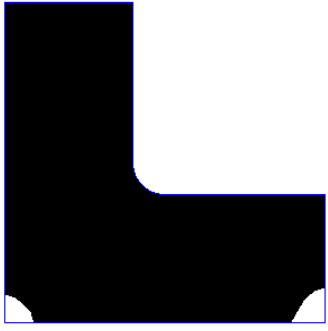

**Figure 3:** Induced topology with 95% volume fraction.

Once a topology $\Omega^\tau$ of desired volume fraction has been computed, one must repeat the finite element analysis and sensitivity computations, leading to the fixed-point algorithm discussed in [49], [50], [51], consisting of the following three steps: (1) solve the finite element problem over $\Omega^\tau$ (2) re-compute the topological sensitivity, and (3) find a new value of $\tau$ for the desired volume fraction as shown in Figure 4.

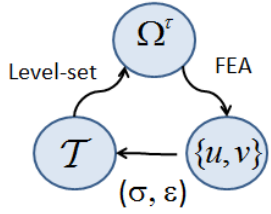

**Figure 4:** Fixed point iteration involving three quantities.

One important feature of the fixed-point iteration is that it allows for the reintroduction of previously deleted material within design domain. This feature is explained in [52], and is illustrated through Figure 5 where a cantilever beam is optimized. It can be observed that material removed at one volume fraction (marked in 'box') is recovered at a later step.

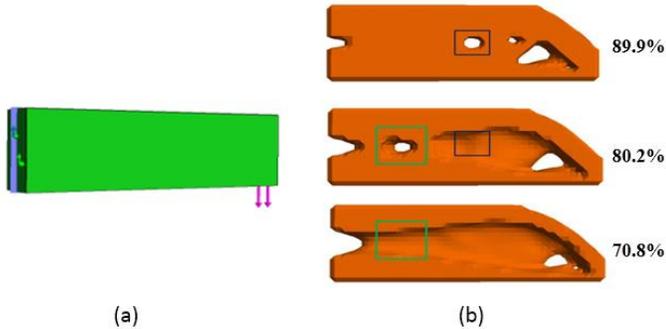

**Figure 5:** (a) A typical cantilever beam problem; (b) Evolving topologies with different volume fractions.

The fixed-point iteration, at a given volume fraction, is repeated (in typically 2~3 iterations), until the compliance converges. Then, the target volume fraction is decreased until the desired volume is reached, resulting in a series of Pareto-optimal topologies. This concept can be easily generalized to 3D [53], see Figure 6 for example illustrating the Pareto-optimal for a 3D structural problem.

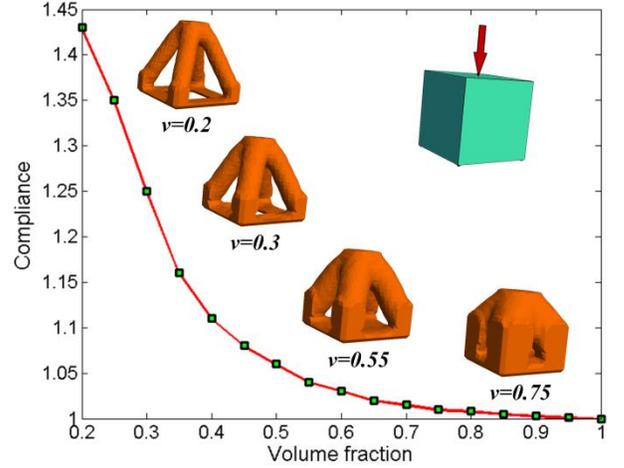

**Figure 6:** Pareto-optimal curve and optimal topologies of a 3D cube.

### 3.3 Constraints implementation

We now consider the constraints. Specifically, consider the thermo-elastic TO problem posed earlier in Equation (2.13). The constraints can be combined with the objective function to define the augmented Lagrangian [54]:

$$L(\mathbf{d},\Omega;\gamma_i,\mu_i) \equiv \varphi + \sum_{i=1}^{m} \bar{L}_i(\mathbf{d},\Omega;\gamma_i,\mu_i) \qquad (3.39)$$

where

$$\bar{L}_i(\mathbf{d},\Omega;\gamma_i,\mu_i) = \begin{cases} \mu_i g_i + \frac{1}{2}\gamma_i(g_i)^2 & \mu_i + \gamma_i g_i > 0 \\ \frac{1}{2}\mu_i^2/\gamma_i & \mu_i + \gamma_i g_i \leq 0 \end{cases} \qquad (3.40)$$

where

$L$ : Augmented Lagrangian

$\bar{L}_i$ : Auxiliary Lagrangian

$\mu_i$ : Lagrangian multipliers  (3.41)

$\gamma_i$ : Penalty parameters

By using the sensitivity definition in Equation (3.1), the gradient of augmented Lagrangian is given by:

$$L' = \varphi' + \sum_{i=1}^{m} \bar{L}_i' \qquad (3.42)$$

where

$$\bar{L}_i' = \begin{cases} (\mu_i + \gamma_i g_i) g_i' & \mu_i + \gamma_i g_i > 0 \\ 0 & \mu_i + \gamma_i g_i \leq 0 \end{cases} \qquad (3.43)$$

The sensitivity of the objective and each of the constraints can be computed using Equation (3.12). The Lagrangian multipliers and penalty parameters are initialized to an arbitrary set of positive values. Then the augmented Lagrangian is minimized using, for example, conjugate gradient method. In every iteration, the Lagrangian multipliers are updated as follows:

$$\mu_i^{k+1} = \max\{\mu_i^k + \gamma_i g_i(\hat{x}^k), 0\}, i = 1, 2, 3, ..., m \quad (3.44)$$

where the $\hat{x}^k$ is the local minimum at the current $k$ iteration.

The penalty parameters are updated via:

$$\gamma_i^{k+1} = \begin{cases} \gamma_i^k & \min(g_i^{k+1}, 0) \le \varsigma \min(g_i^k, 0) \\ \max(\eta \gamma_i^k, k^2) & \min(g_i^{k+1}, 0) > \varsigma \min(g_i^k, 0) \end{cases} \quad (3.45)$$

where $0 < \varsigma < 1$ and $\eta > 0$, $\varsigma = 0.25$ and $\eta = 10$ [54]. Readers are referred to [55] for details.

### 3.4 Proposed algorithm

Finally, the proposed algorithm proceeds as follows (see Figure 7):

1. The domain is discretized using finite elements (here 3D hexahedral elements). The optimization starts at a volume fraction of 1.0. The 'volume decrement' $\Delta v$ is set to 0.025. The initial values of Lagrangian multiplier and penalty number are set as 100 and 10.
2. The thermal problem (if necessary) and the structural problem in Equations (2.1) and (2.2) are solved.
3. The constraint values are calculated, and the optimization parameters (multiplier and penalty) are updated.
4. If any of the constraints is violated, the algorithm proceeds to step-9, else, it proceeds to step-5.
5. The sensitivities are calculated for each of the constraints, and the augmented element sensitivity field is computed.
6. Treating the augmented sensitivity field as a level-set; a new topology with a volume fraction of ($v - \Delta v$) is extracted.
7. The compliance is now computed over the new topology. If the compliance has converged, then the optimization moves to the next step, else it goes to step 9.
8. The current volume fraction is set to ($v - \Delta v$). If the target volume fraction has not been reached, the optimization returns to step-2 to repeat iterations; else, terminate iteration and exit.
9. Step-size is reduced; check if volume decrement is below threshold. If not, the optimization returns to step-2; else, terminate the iteration.

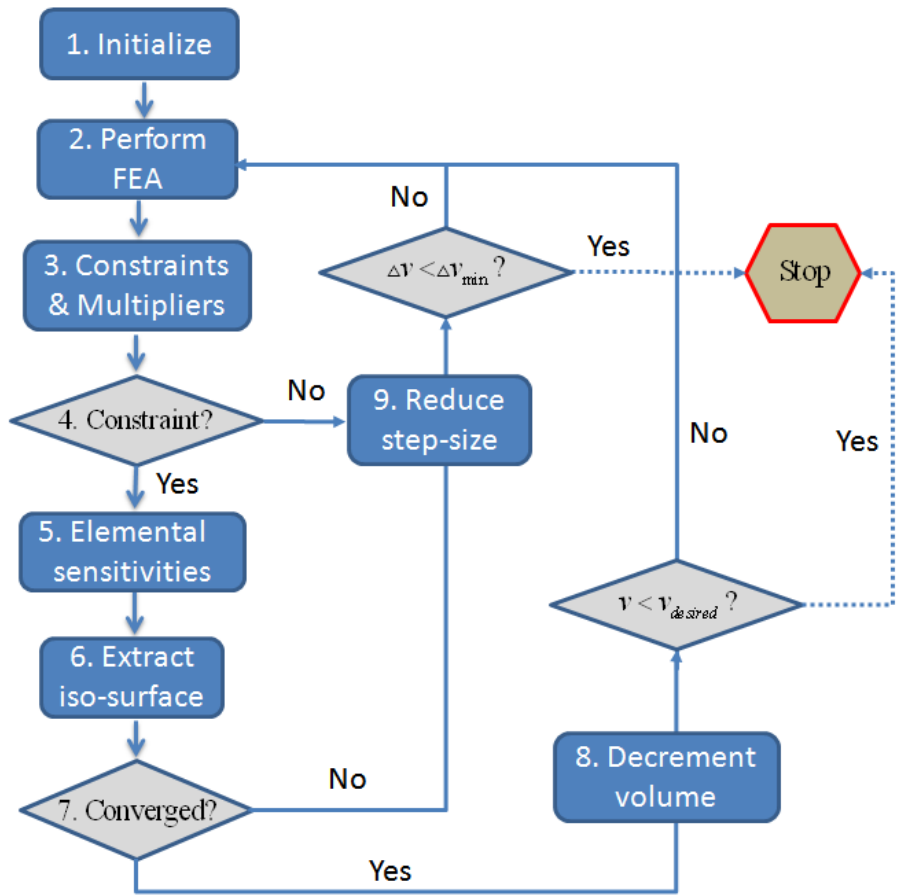

**Figure 7:** An overview of our algorithm.

## 4. Numerical experiments

In this Section, we demonstrate the efficacy of the proposed algorithm through numerical experiments. The default parameters are as follows:

- The material is assumed to be steel, i.e., the elastic modulus is $E = 2e11\ Pa$, the Poisson ratio is $\nu = 0.3$ and the coefficient of thermal expansion $\alpha = 1.2e-5/\ ^oC$.
- The reference temperature is $23^0C$.
- Unless otherwise noted, the p-norm value used for computing the p-norm stress is 6.
- 8-noded hexahedral elements are used for 3D FEA.
- All experiments were conducted using C++ on a Windows 7 64-bit machine with the following hardware: Intel I7 960 CPU quad-core running at 3.2GHz with 6 GB of memory.
- The desired volume fraction is 0.25, unless otherwise noted. In other words, the optimization terminates if the constraints are violated or if the desired volume fraction of 0.25 is reached.

The numerical experiments are organized as follows. Section 5.1 is a benchmark example to study the effectiveness of the proposed method for uniformly elevated temperature; both compliance and stress dominated problems are considered. In Section 5.2, another benchmark example is considered to study the effect of spatially varying temperature. In Section 5.3, a case-study involving a flange subject to a uniform temperature increase is considered. Finally, in Section 5.4, a case study is considered where the structure is subject to temperature gradient fields. Important conclusions are drawn for each of the examples.

### 4.1 Benchmark: clamped beam with a point load

The aim of this experiment is two-fold: (1) illustrate the proposed algorithm for a benchmark problem [12], (2) show the impact of temperature variations on the final topology.

The structure is illustrated in Figure 8 [12], units are in meters, the load is $10^5$ N, the thickness is 0.02m, and the structure is also subject to a homogeneous temperature increase, specified below. Since the thickness is small, the problem can be modeled as plane-stress [12]. However, it is modeled here in 3D, and the domain is meshed with 15,000 hexahedral elements.

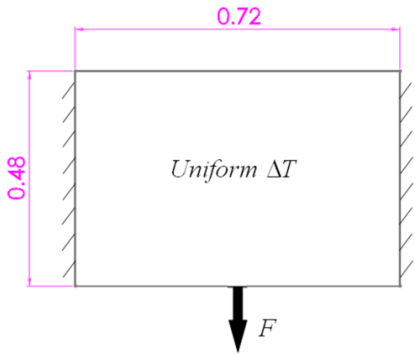

**Figure 8:** The bi-clamped structure with a central point load.

Compliance formulation (stiff designs)

We first consider compliance-constrained thermo-elastic TO problem:

$$\underset{\Omega \subset \Psi}{Min}\ |\Omega|$$
$$|\Omega| \geq 0.25 |\Psi|$$
$$J \leq 5J_0 \tag{4.1}$$
subject to
$$\mathbf{Kd} = \mathbf{f}_{st} + \mathbf{f}_{th}$$
$$\Delta t:\ \text{Specified}$$

Observe that the thermal problem in Equation (2.1) need not be considered since the temperature increase is prescribed.

If the temperature increase $\Delta t$ is $1^0C$, the optimized topology for a 0.25 volume fraction is illustrated in Figure 9 where the final compliance and stress are 1.94 and 1.02 times their initial values, respectively. The computational time is 58 seconds, involving 242 FEAs; the topology is identical to the one obtained in [12]. The final compliance of the structure with 25% volume fraction is almost twice the initial compliance of the structure with 100% volume fraction, while the stress has not increased significantly.

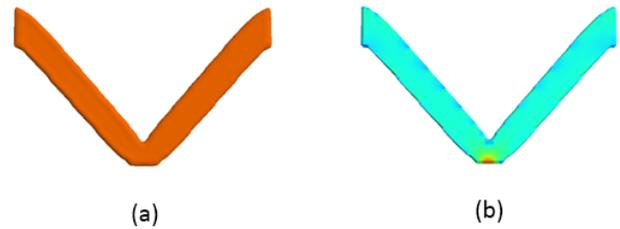

**Figure 9:** (a) Optimized topology; (b) Stress distribution of the compliance-constrained TO.

Next, we consider the impact of temperature change on the final topology. The target volume fraction was set to 0.25 and the final topologies are illustrated in Figure 10 for a temperature change ranging from $-5^oC$ to $+5^oC$. As one can observe, the final topology is a strong function of the temperature change, especially for a positive change.

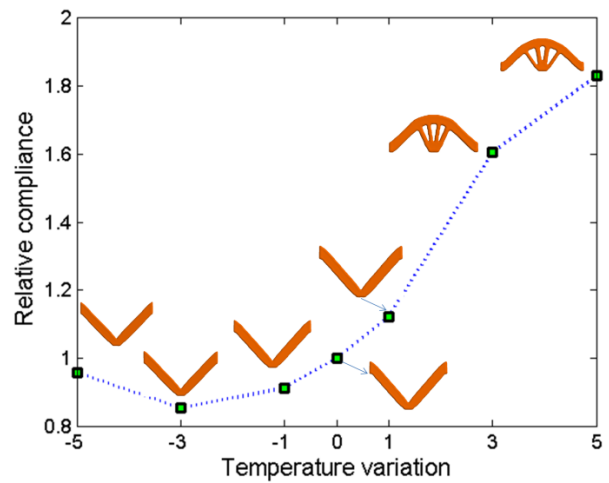

**Figure 10:** The final topologies for different temperature variations for problem in Equation (4.1).

Observe from Figure 10 that if the temperature is increased, the compliance monotonically increases; on the other hand, if the temperature is decreased, the compliance first decreases, and then increases. One possible reason is that when the temperature decrement is small (e.g., $-1^0C$), the compressive thermal load partly cancels the tensile structural load, resulting in a smaller compliance value.

The relative magnitudes of thermal and mechanical loads are summarized in Table 1. By comparing the cases with temperature variations of $\Delta t = 5^oC$ and $\Delta t = -5^oC$ which have close magnitudes of thermal loads, the influence of thermally induced expansion and contraction on final topologies can be clearly seen.

**Table 1:** Load ratios for different temperature variations

| $\Delta t$ ($^0C$) | -5 | -3 | -1 | 0 | 1 | 3 | 5 |
|---|---|---|---|---|---|---|---|
| $\|f_{th}\|/\|f_{st}\|$ | 3.83 | 2.06 | 0.35 | 0 | 0.42 | 2.08 | 3.91 |

Stress formulation (strong designs)

We pose a stress dominated thermo-elastic TO as follows:

$$\underset{\Omega \subset \Psi}{Min} |\Omega|$$
$$|\Omega| \geq 0.25|\Psi|$$
$$\sigma \leq 2\sigma_0 \qquad (4.2)$$
subject to
$$\mathbf{Kd} = \mathbf{f}_{st} + \mathbf{f}_{th}$$
$\Delta t$ : Specified

Similar to the previous experiment, the temperature is uniformly elevated by $1^0C$. The resulting topology with the 0.25 volume fraction is illustrated in Figure 11 where its final compliance and stress equal to 2.08 and 0.97 times their initial values, respectively. The computing time was 122 seconds involving 363 FEA. The increased computing time is due to the additional adjoint FEA that needs to be performed.

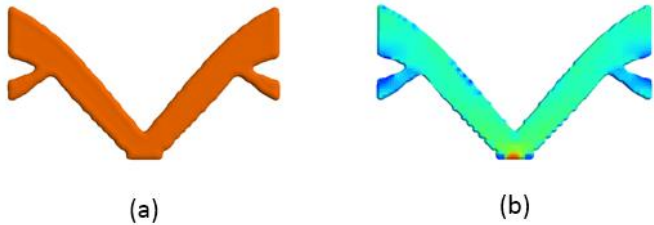

**Figure 11:** (a) Final optimized topology; (b) Stress distribution of the stress-constrained TO.

Comparing Figure 11 and Figure 9, it can be observed that: compliance and stress dominated TO lead to slightly different topologies, and the topology in Figure 9 has lower compliance while the topology in Figure 11 and has lower stress, as expected.

The final topologies for different temperature variations are illustrated in Figure 12. As one can observe, the topologies are significantly different from those in Figure 10.

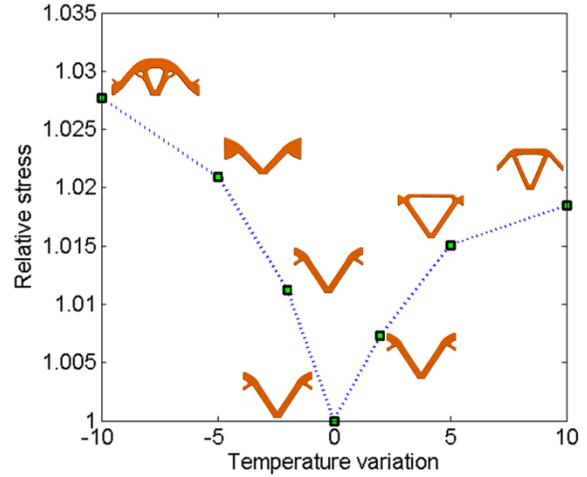

**Figure 12:** The final topologies for different temperature variations.

### 4.2 Benchmark: clamped beam with distributed loads

The aim of this experiment is to study the impact of non-uniform temperature on the final topology.

We once again consider the bi-clamped beam but with a distributed load as illustrated in Figure 13 [12]. The dimension of this beam is $0.5m \times 0.28m \times 0.01m$ and the distributed load is $P = 6e5 Pa$. Again, the problem is modeled in 3D, and the domain is meshed with 15,000 hexahedral elements.

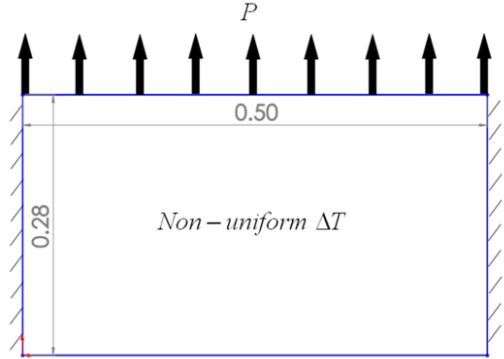

**Figure 13:** The bi-clamped structure with a distributed load.

Compliance formulation (stiff designs)

The specific problem being considered here is:

$$\underset{\Omega \subset \Psi}{Min} |\Omega|$$
$$|\Omega| \geq 0.30|\Psi|$$
$$J \leq 5J_0 \qquad (4.3)$$
subject to
$$\mathbf{Kd} = \mathbf{f}_{st} + \mathbf{f}_{th}$$
$$\mathbf{K}_t \mathbf{t} = \mathbf{q}$$

If the temperature is uniformly elevated by $20^0 C$, the resulting topology and stress distribution are illustrated in Figure 14. This is consistent with the results in [12]. The resulting compliance and stress values are 3.62 and 1.97 times corresponding initial values.

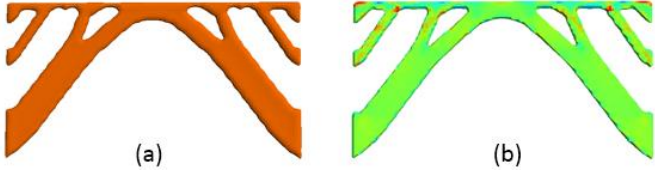

**Figure 14:** Final topology (a) and stress distribution (b) when the structure in Figure 13 is subject to uniform temperature rise.

Next, we consider the impact of spatially varying temperature on the optimal designs. Specifically, we increased the temperature on the left edge by $0^0 C$, and on the right edge by $40^0 C$, i.e., the average change in temperature is $20^0 C$. The final topology and its stress distribution are shown in Figure 15 where the asymmetry is due to the spatial thermal gradient. Comparison between Figure 14 and Figure 15 highlights the importance of accounting for spatially distributed temperature profiles.

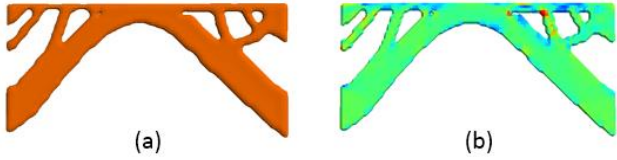

**Figure 15:** Final topology (a) and stress distribution (b) when the structure is subject to a spatially temperature gradient with resulting compliance and stress are as large as 4.17 and 1.98 times their initial values.

Stress formulation (strong designs)

Next a stress dominated problem is considered for the above problem in Figure 13:

$$\underset{\Omega \subset \Psi}{Min} |\Omega|$$
$$|\Omega| \geq 0.30 |\Psi|$$
$$\sigma \leq 2\sigma_0 \quad (4.4)$$
subject to
$$\mathbf{Kd} = \mathbf{f}_{st} + \mathbf{f}_{th}$$
$$\mathbf{K}_t \mathbf{t} = \mathbf{q}$$

On the left edge there was no temperature change, i.e., $\Delta t = 0^0 C$, and on the right edge the change was $\Delta t = 40^0 C$. The final topology and stress distribution results are illustrated in Figure 16. Their final compliance and stress are 4.97 and 1.84 times their initial values. Observe the strong asymmetry in the stress-dominated problem.

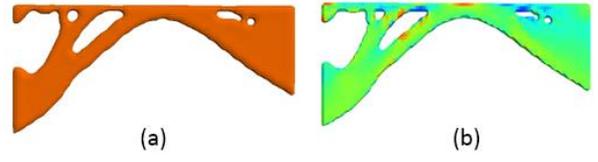

**Figure 16:** Final topology (a) and stress distribution (b) of stress-constrained TO subject to spatially temperature gradient.

Comparing the results in Figure 16 with Figure 15, it is clear: firstly, the two optimization problems lead to distinct topologies; also, at the same final volume fraction, a compliance minimization leads to a lower compliance result while a stress minimization leads to a lower stress value.

### 4.3 Case study: flange

The purpose of this section is to show the robustness of the proposed algorithm for a non-trivial application. In particular, a thermo-elastic TO problem over a flange is studied in this section. Flanges are commonly used, for example, to fasten pipes and rail-joints, and they are often subject to temperature changes. The dimensions of the flange and boundary conditions are illustrated in Figure 17. The flange is fixed at the two bolt centers, and a vertical force of $10^5 N$ is applied as shown. For FEA, 51,500 hexahedral elements are used to discretize the design domain, resulting in 175,374 DOF.

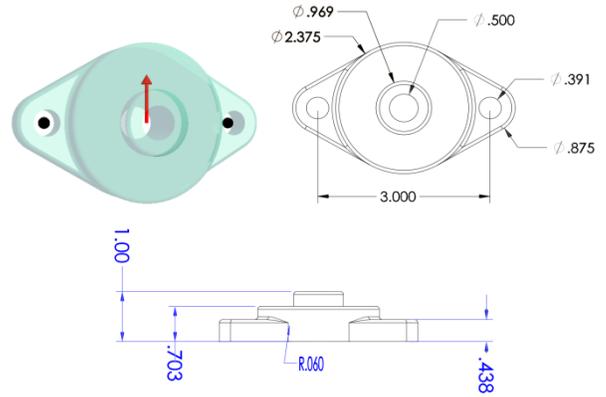

**Figure 17:** Flange structure and dimensions (unit: meter).

The specific thermo-elastic TO problem considered here is:

$$\underset{\Omega \subset \Psi}{Min} |\Omega|$$
$$|\Omega| \geq 0.25 |\Psi|$$
$$J \leq 5 J_0 \quad (4.5)$$
$$\sigma \leq 1.5 \sigma_0$$
subject to
$$\mathbf{Kd} = \mathbf{f}_{st} + \mathbf{f}_{th}$$

First, a pure elastic problem (i.e., zero thermal load in Equation (4.5)) is considered. The resulting topology is illustrated in Figure 18 and the final constraint values are shown in Table 2 where the optimization terminates due to the active stress constraint identified with a "box".

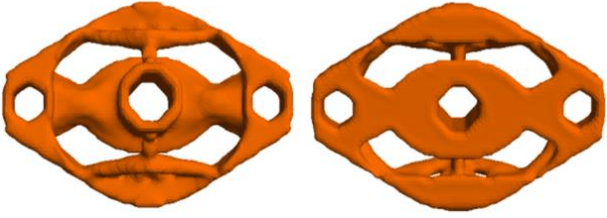

**Figure 18:** Top view and bottom view of final topology for the pure elastic flange problem.

**Table 2:** Constraints and results for problem in Figure 18.

| Initial Constraints | Final Results | Volume and time (sec) | Final load ratio |
|---|---|---|---|
| $J \leq 5J_0$ | $J = 4.89J_0$ | $V = 0.36$ | $\frac{\|f_{th}\|}{\|f_m\|} = 0$ |
| $\sigma \leq 1.5\sigma_0$ | $\boxed{\sigma = 1.50\sigma_0}$ | $T = 212.39$ | |

Then, the thermal effect is added; we make the structure subject to a uniform temperature elevation of 30°C.

The optimized topology, computed in 160 FEAs, is illustrated in Figure 19. Other results are summarized in Table 3; this problem terminated due to an active compliance constraint. Although the thermal load is small compared to the structural load, as noted in the fourth column of Table 3, this has a significant effect on the final topology.

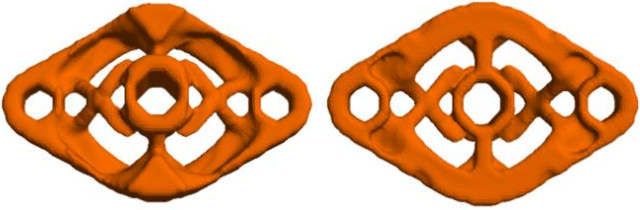

**Figure 19:** Top view and bottom view of final topology of the flange subject to a uniform temperature rise.

**Table 3:** Constraints and results for problem in Equation (4.5).

| Initial Constraints | Final Results | Final volume & time (sec) | Final load ratio |
|---|---|---|---|
| $J \leq 5J_0$ | $\boxed{J = 5J_0}$ | $V = 0.44$ | $\frac{\|f_{th}\|}{\|f_m\|} = 0.04$ |
| $\sigma \leq 1.5\sigma_0$ | $\sigma = 1.48\sigma_0$ | $T = 239.98$ | |

### 4.4 Case study: exhaust system

Next we consider engine exhaust-washed structure, used in a low observable supersonic aircraft; this was first studied by J. Deaton [17]. Due to low radar observability requirement, engine and exhaust system are buried inside the aircraft. Because of the space restriction, the exhaust system is fixed onto the aircraft skins; thermal expansion is therefore limited. In order to reduce infrared detectability, hot exhaust gas is cooled within the exhaust duct.

A simplified exhaust system conception with its dimensions (unit: meter) is illustrated in Figure 20 where the structure is fixed at left and right ends, and fixtures. A temperature at intake is assumed $400^0 C$ and cooled down to $100^0 C$ at output nozzle. For FEA, the domain is meshed with 54,080 hexahedral elements, resulting in 208,374 DOF.

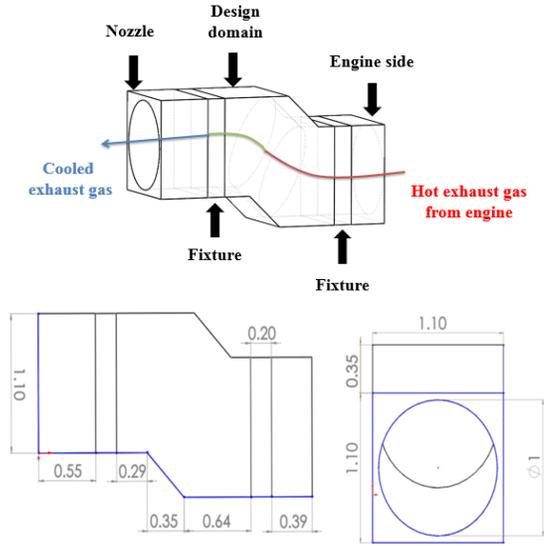

**Figure 20:** Conceptual exhaust system and its dimensions.

The specific thermo-elastic TO problem solved here is:

$$\underset{\Omega \subset \Psi}{Min} |\Omega|$$
$$J \leq 1.5 J_0$$
$$\sigma \leq 1.5 \sigma_0 \qquad (4.6)$$
subject to
$$\mathbf{Kd} = \mathbf{f}_{st} + \mathbf{f}_{th}$$
$$\mathbf{K}_t \mathbf{t} = \mathbf{q}$$

The final topology is illustrated in Figure 21. Optimization results are summarized in Table 4. On termination, the compliance constraint is active and the maximum p-norm stress is reduced.

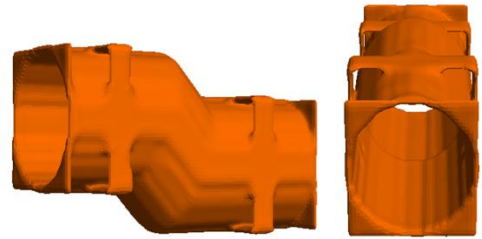

**Figure 21**: Side view and front view of the optimized exhaust

**Table 4:** Constraints and results for problem in Equation (4.6).

| Initial Constraints | Final Results | Final volume & time (s) |
|---|---|---|
| $J \leq 1.5 J_0$ | $\boxed{J = 1.50 J_0}$ | $V = 0.46$ |
| $\sigma \leq 1.5\sigma_0$ | $\sigma = 0.83\sigma_0$ | $T = 531.4$ |

## 5. Conclusions

The main contribution of this paper is a comprehensive method for solving constrained thermo-elastic TO problems. For thermal scenario with complex temperature fields, the sensitivity to compliance and p-norm stress are derived. Augmented Lagrangian method was used for multi-constrained thermo-elastic TO. The assembly-free FEA method was implemented for acceleration.

As the numerical experiments reveal, the impact of both uniform temperature variations and spatially thermal gradients on the final topologies can be significant for certain problems. Future work will focus on including other constraints including eigen-modes and large-deformation buckling in the thermo-elastic TO analysis.

## Acknowledgements

The authors would like to thank the support of National Science Foundation through grants CMMI-1232508, CMMI-1161474, and IIP-1500205.

## Appendix

In order to prove the correctness of sensitivity analysis in this paper, the displacement sensitivity in Equation (3.23) is compared with its counterpart in [44]. As shown in Chapter 2.8.1 of [44] the sensitivity of a weakly-coupled thermo-elastic problem can be derived as below.

The finite element equations are given for the two systems:

$$\mathbf{H}\Delta\mathbf{t} = \mathbf{f}_{th} \quad (1)$$

$$\mathbf{Kd} = \mathbf{f} \quad (2)$$

where

$$\Delta\mathbf{t} = \mathbf{t} - \mathbf{t}_0 \quad (3)$$

The force vector on the right-hand side of Equation (2) is the sum of the thermal load and design-independent mechanical load:

$$\mathbf{f} = \mathbf{f}_{th} + \mathbf{f}_{st} \quad (4)$$

If we have an interests in the displacement at a given point $a$. Using adjoint methods, the equation can be formed as following:

$$d_a = \mathbf{1}^T\mathbf{d} + \boldsymbol{\lambda}_1^T(\mathbf{H}\Delta\mathbf{t} - \mathbf{f}_{th}) + \boldsymbol{\lambda}_2^T(\mathbf{Kd} - \mathbf{f}) \quad (5)$$

Then, the sensitivity with respect to element pseudo-density can be shown as:

$$d_a' = \boldsymbol{\lambda}_1^T(\mathbf{H}'\Delta\mathbf{t} - \mathbf{f}_{th}') + \boldsymbol{\lambda}_2^T(\mathbf{K}'\mathbf{d} - \mathbf{f}') \quad (6)$$

where the two adjoints are defined as:

$$\mathbf{K}\boldsymbol{\lambda}_2 = -\mathbf{1} \quad (7)$$

$$\mathbf{H}\boldsymbol{\lambda}_1 = \boldsymbol{\lambda}_2^T\frac{\partial\mathbf{f}}{\partial\mathbf{t}} = -(\mathbf{K}^{-1}\mathbf{1})^T\mathbf{H} \quad (8)$$

Plugging the two adjoints into Equation (6):

$$\begin{aligned} d_a' &= \boldsymbol{\lambda}_1^T(\mathbf{H}'\Delta\mathbf{t} - \mathbf{f}_{th}') + \boldsymbol{\lambda}_2^T(\mathbf{K}'\mathbf{d} - \mathbf{f}') \\ &= (\mathbf{H}^T\mathbf{K}^{-1}\mathbf{1} + \mathbf{1}^T\mathbf{K}^{-1}\mathbf{H})\mathbf{t}' + (\mathbf{K}^{-1}\mathbf{1})^T(\mathbf{H}'\Delta\mathbf{t} - \mathbf{K}'\mathbf{d}) \end{aligned} \quad (9)$$

where Equation (9) is shown identical to Equation (3.23).

## References

[1] H. A. Eschenauer and N. Olhoff, "Topology optimization of continuum structures: A review," *Applied Mechanics Review*, vol. 54, no. 4, pp. 331–389, 2001.

[2] G. I. N. Rozvany, "A critical review of established methods of structural topology optimization," *Structural and Multidisciplinary Optimization*, vol. 37, no. 3, pp. 217–237, 2009.

[3] E. Kesseler, "Multidisciplinary design analysis and multi-objective optimisation applied to aircraft wing," *WSEAS transactions on systems and Control and Cybernetics*, vol. 1, no. 2, p. 221 227, 2006.

[4] J. J. Alonso, "Aircraft design optimization," *Mathematics and Computers in Simulation*, vol. 79, no. 6, pp. 1948–1958, 2009.

[5] V. H. Coverstone-Carroll, "Optimal multi-objective low-thrust spacecraft trajectories," *Comput. Methods Appl. Mech. Eng.*, vol. 186, pp. 387–402, 2000.

[6] L. Wang, "Automobile body reinforcement by finite element optimization," *Finite Elements in Analysis and Design*, vol. 40, no. 8, pp. 879–893, 2004.

[7] L. Harzheim, "A review of optimization of cast parts using topology optimization II-Topology optimization with manufacturing constraints," *Structural and Multidisciplinary Optimization*, vol. 31, no. 5, pp. 388–299, 2006.

[8] G. K. Ananthasuresh, S. Kota, and Y. Gianchandani, "A methodical approach to the design of compliant micromechanisms," in *Solid State Sensor and Actuator Workshop*, 1994, pp. 189–192.

[9] S. Nishiwaki, "Topology Optimization of Compliant Mechanisms using the Homogenization Method," *International Journal for Numerical Methods in Engineering*, vol. 42, pp. 535–559, 1998.

[10] T. E. Bruns and D. A. Tortorelli, "Topology optimization of non-linear elastic structures and compliant mechanisms," *Computer Methods in Applied Mechanics and Engineering*, vol. 190, no. 26–27, pp. 3443–3459, 2001.

[11] Z. Luo, "Compliant mechanism design using multi-objective topology optimization scheme of continuum structures," *Structural and Multidisciplinary Optimization*, vol. 30, pp. 142–154, 2005.

[12] H. Rodrigues and H. Fernades, "A material based model for topology optimization of thermoelastic structures," *International Journal for Numerical Methods in Engineering*, vol. 38, pp. 1951–65, 1995.

[13] O. Sigmund and S. Torquato, "Design of materials with extreme thermal expansion using a three-phase topology optimization method," *Journal of the Mechanics and Physics of Solid*, vol. 45, pp. 1037–1067, 1997.

[14] C. Jog, "Distributed-parameter optimization and topology design for non-linear thermoelasticity," *Computer Methods in Applied Mechanics and Engineering*, vol. 132, no. 1, pp. 117–134, 1996.

[15] O. Sigmund, "A 99 line topology optimization code written in Matlab," *Structural and Multidisciplinary Optimization*, vol. 21, no. 2, pp. 120–127, 2001.

[16] D. Li and X. Zhang, "Topology Optimization of Thermo-Mechanical Continuum Structure," presented at the 2010 IEEE/ASME International Conference on Advanced Intelligent Mechatronics, Canada, 2010.

[17] J. Deaton and R. V. Grandhi, "Stiffening of Thermally Restrained Structures via Thermoelastic Topology Optimization," presented at the 53rd AIAA/ ASME/ ASCE/ AHS/ ASC Structures, Structural Dynamics and Materials Conference, Honolulu, Hawaii, 2012.

[18] X. Yang and Y. Li, "Structural topology optimization on dynamic compliance at resonance frequency in thermal environments," *Struct Multidisc Optim*, vol. 49, no. 1, pp. 81–91, Jul. 2013.

[19] X. Liu, C. Wang, and Y. Zhou, "Topology optimization of thermoelastic structures using the guide-weight method," *Sci. China Technol. Sci.*, vol. 57, no. 5, pp. 968–979, May 2014.

[20] W. Zhang, J. Yang, Y. Xu, and T. Gao, "Topology optimization of thermoelastic structures: mean compliance minimization or elastic strain energy minimization," *Struct Multidisc Optim*, vol. 49, no. 3, pp. 417–429, Sep. 2013.

[21] Y. Chen, S. Zhou, and Q. Li, "Multiobjective topology optimization for finite periodic structures," *Computers & Structures*, vol. 88, no. 11–12, pp. 806–811, 2010.

[22] P. Pedersen and N. L. Pedersen, "Strength optimized designs of thermoelastic structures," *Struct Multidisc Optim*, vol. 42, no. 5, pp. 681–691, Jul. 2010.

[23] M. Stolpe and K. Svanberg, "An alternative interpolation scheme for minimum compliance topology optimization," *Structural and Multidisciplinary Optimization*, vol. 22, pp. 116–124, 2001.

[24] T. Gao and W. Zhang, "Topology optimization involving thermo-elastic stress loads," *Structural and Multidisciplinary Optimization*, vol. 42, pp. 725–738, 2010.

[25] T. Gao and W. Zhang, "Topology optimization involving thermo-elastic stress loads," *Struct Multidisc Optim*, vol. 42, no. 5, pp. 725–738, Jun. 2010.

[26] P. Pedersen and N. L. Pedersen, "Interpolation/penalization applied for strength design of 3D thermoelastic structures," *Struct Multidisc Optim*, vol. 45, no. 6, pp. 773–786, Jan. 2012.

[27] D. J. Munk, G. A. Vio, and G. P. Steven, "Topology and shape optimization methods using evolutionary algorithms: a review," *Struct Multidisc Optim*, pp. 1–19, May 2015.

[28] X. Huang and Y. M. Xie, "A new look at ESO and BESO optimization methods," *Structural and Multidisciplinary Optimization*, vol. 35, no. 1, pp. 89–92, 2008.

[29] Q. Li, G. P. Steven, and Y. M. Xie, "Displacement minimization of thermoelastic structures by evolutionary thickness design," *Computer Methods in Applied Mechanics and Engineering*, vol. 179, no. 3–4, pp. 361–378, Sep. 1999.

[30] Q. Li, G. P. Steven, O. M. Querin, and Y. M. Xie, "Structural topology design with multiple thermal criteria," *Engineering Computations*, vol. 17, no. 6, pp. 715–734, 2000.

[31] Y. M. X. Qing Li Grant P.Steven, "THERMOELASTIC TOPOLOGY OPTIMIZATION FOR PROBLEMS WITH VARYING TEMPERATURE FIELDS," *Journal of Thermal Stresses*, vol. 24, no. 4, pp. 347–366, 2001.

[32] S. Osher and J. A. Sethian, "Fronts Propagating with Curvature-dependent Speed: Algorithms Based on Hamilton-Jacobi Formulations," *J. Comput. Phys.*, vol. 79, no. 1, pp. 12–49, Nov. 1988.

[33] J. Sethian, "Structural boundary design via level set and immersed interface methods," *Journal of Computational Physics*, vol. 163, no. 2, pp. 489–528, 2000.

[34] Q. Xia and M. Y. Wang, "Topology Optimization of Thermoelastic Structures Using Level Set Method," presented at the EngOpt 2008 - International Conference on Engineering Optimization, Rio de Janeiro, Brazil, 2008.

[35] N. Vermaak, G. Michailidis, G. Parry, R. Estevez, G. Allaire, and Y. Bréchet, "Material interface effects on the topology optimization of multi-phase structures using a level set method," *Struct Multidisc Optim*, pp. 1–22, Jun. 2014.

[36] S. Deng, K. Suresh, and J. Joo, "Stress-Constrained Thermo-Elastic Topology Optimization: A Topological Sensitivity Approach," presented at the Proceedings of the ASME IDETC/CIE Conference, Buffalo, NY, USA, 2014.

[37] S. Deng and K. Suresh, "TOPOLOGY OPTIMIZATION UNDER LINEAR THERMO-ELASTIC BUCKLING," presented at the ASME-IDETC Conference, Charlotte, NC, 2016, vol. DETC2016-59408.

[38] P. Yadav and K. Suresh, "Large Scale Finite Element Analysis Via Assembly-Free Deflated Conjugate Gradient," *J. Comput. Inf. Sci. Eng*, vol. 14, no. 4, pp. 41008-1-9, 2014.

[39] R. B. Hetnarski, J. Ignaczak, N. Noda, N. Sumi, and Y. Tanigawa, *Theory of Elasticity and Thermal Stresses: Explanations, Problems and Solutions*. Springer, 2013.

[40] J. Deaton and R. V. Grandhi, "Topology Optimization of Thermal Structures with Stress Constraints," presented at the 54th AIAA/ASME/ASCE/ASC Structures, Structural Dynamics, and Materials Conference, Boston, MA, 2013.

[41] P. Pedersen and N. L. Pedersen, "Strength optimized designs of thermoelastic structures," *Struct Multidisc Optim*, vol. 42, no. 5, pp. 681–691, Jul. 2010.

[42] M. P. Bendsøe, *Topology optimization theory, methods and applications*. Berlin Heidelberg: Springer Verlag, 2003.

[43] M. Bruyneel and P. Duysinx, "Note on topology optimization of continuum structures including self-weight," *Struct Multidisc Optim*, vol. 29, no. 4, pp. 245–256, Nov. 2004.

[44] M. Bendsøe and O. Sigmund, *Topology Optimization: Theory, Methods and Application*, 2nd ed. Springer, 2003.

[45] K. Suresh and M. Takalloozadeh, "Stress-Constrained Topology Optimization: A Topological Level-Set Approach," *Structural and Multidisciplinary Optimization*, vol. 48, no. 2, pp. 295–309, 2013.

[46] G. Allaire, F. Jouve, and A. M. Toader, "Structural Optimization using Sensitivity Analysis and a Level-set

Method," *Journal of Computational Physics*, vol. 194, no. 1, pp. 363–393, 2004.

[47] S. Amstutz and H. Andra, "A new algorithm for topology optimization using a level-set method," *Journal of Computational Physics*, vol. 216, pp. 573–588, 2006.

[48] K. Suresh, "A 199-line Matlab code for Pareto-optimal tracing in topology optimization," *Structural and Multidisciplinary Optimization*, vol. 42, no. 5, pp. 665–679, 2010.

[49] J. Céa, S. Garreau, P. Guillaume, and M. Masmoudi, "The shape and topological optimization connection," *Computer Methods in Applied Mechanics and Engineering*, vol. 188, no. 4, pp. 713–726, 2000.

[50] J. A. Norato, M. P. Bendsøe, R. B. Haber, and D. A. Tortorelli, "A topological derivative method for topology optimization," *Structural and Multidisciplinary Optimization*, vol. 33, pp. 375–386, 2007.

[51] K. Suresh, "Efficient Generation of Large-Scale Pareto-Optimal Topologies," *Structural and Multidisciplinary Optimization*, vol. 47, no. 1, pp. 49–61, 2013.

[52] A. Krishnakumar and K. Suresh, "Hinge-Free compliant mechanism design via the Topological Level-Set," *Journal of Mechanical Design*, vol. 137, no. 3, 2015.

[53] A. A. Novotny, "Topological-Shape Sensitivity Method: Theory and Applications," *Solid Mechanics and its Applications*, vol. 137, pp. 469–478, 2006.

[54] J. Nocedal and S. Wright, *Numerical Optimization*. Springer, 1999.

[55] S. Deng and K. Suresh, "Multi-constrained topology optimization via the topological sensitivity," *Structural and Multidisciplinary Optimization*, vol. 51, no. 5, pp. 987–1001, 2015.